# REDUCTION OF MONITORING REGISTERS ON SOFTWARE DEFINED NETWORKS

Luz Angela Aristizábal Q.[1] and Nicolás Toro G.[2]

[1]Department of Computation, Faculty of Management, Universidad Nacional de Colombia.
[2]Department of Electrical and Electronic Engineering, Universidad Nacional de Colombia.

*ABSTRACT*

*Characterization of data network monitoring registers allows for reductions in the number of data, which is essential when the information flow is high, and implementation of processes with short response times, such as interchange of control information between devices and anomaly detection is required. The present investigation applied wavelet transforms, so as to characterize the statistic monitoring register of a software-defined network. Its main contribution lies in the obtention of a record that, although reduced, retains detailed, essential information for the correct application of anomaly detectors.*

## 1. INTRODUCTION

Monitoring the activity of a data network involves the verification of operation limits, such as changes in bandwidth that ensure Quality of Service (QoS), server and interconnection device congestion levels, safety conditions, etc. Said verification is a perpetual activity that requires the analysis of large volumes of information, as well as short response times, and must compensate the overhead in both measurement and accuracy [1].

Traffic measurement approaches may be either active or passive. Passive methods take their measurements from traffic that passes through network devices without introducing overhead. Active methods, on the other hand, add traffic to a network by sending packets that are used to obtain network parameters, for example, the latency of a link or a given device [2].

With the emergence of software-defined networks in 2008, a new prospect for the implementation of network monitors was visualized. In this operation model, each switch connected to a controller also includes the generation of statistics associated with the flows of data circulating through its ports [3] [4]. This makes the implementation of passive monitoring simpler, although it causes difficulty in the analysis of large volumes of information [5]. Subsampling techniques have been used in passive monitoring, in order to reduce the amount of data in analysis processes [6][8]. However, in order for an abnormality detector to achieve high accuracy, it is necessary to consider all of the statistics that the network can provide, which can be achieved by considering all statistical information that the software defined network switches can send to the controller.

Data reduction or compression applications generally involve two processes: *data characterization* and coding. The objective of data characterization is to extract the most relevant

International Journal of Computer Science & Information Technology (IJCSIT) Vol 11, No 2, April 2019

information from the original data. Examples include transformations, such as the Wavelet and Fourier transforms. Coding reduces the number of bits necessary to represent data. Coding methods include Shannon-Fano, arithmetic coding, and the Huffman method, which generate probability distributions in order to assign the smallest code to the symbol with the greatest probability of occurrence in the data, and the code with the highest number of bits to the signal data with the lowest probability of occurrence [9].

Wavelet transform has been widely used for the *characterization* of different types of data, which encompass: biological data [10][11], images, and sound and network traffic measurements [12]. However, its effectiveness in data network anomaly detection has not been explicitly proven. There are several investigations regarding software defined networks that have utilized algorithms such as the K-SVD and Orthogonal Matching Pursuit (OMP) methods[13] for traffic parameter reduction, but these emphasize the compression of control plane information.

The interest of the present study was to consider wavelet transform a characterization method that could reduce the statistical data number generated in the network, thus retaining the level of detail that a detector necessitates the evaluation of anomalous behaviours. Given that anomaly detectors use traffic monitor output data as input for their processes, it must be ensured that anomaly detectors generate the same output, both when fed with original data and when fed with transformed data.

Results were evaluated through consideration of the data reduction percentage and contrasting the reliability of the anomaly detector with original vs. reduced data.

This article begins by illustrating the structure of software-defined networks and the statistics obtained by network switches. It continues with the application of the wavelet transform in the characterization process. The method implemented for anomaly detection follows, and experimental results conclude the content of the present document.

## 2. SOFTWARE-DEFINED NETWORKS

### 2.1 STRUCTURE

A Software-defined Networks (SDN) consist of at least two parts: controllers and interconnection devices, such as switches and routers, as shown in Figure 1.

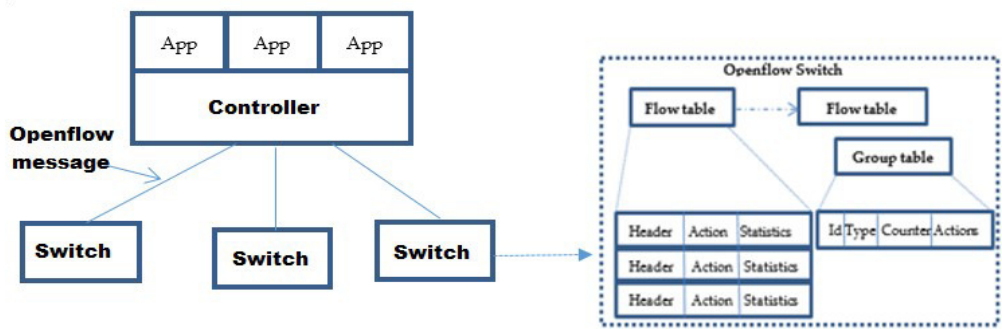

Figure 1. Software-defined Network

The switches, essentially, forward data in accordance with the forwarding policies employed by the controller. A controller is a centralized software that directs the switch's activities with incoming data flows, determines which port will forward data, and whether it is necessary to duplicate or drop said data. This decision is made with each data flow that arrives to the switch. Communication between switches and controllers is established by the "OpenFlow" protocol [4]. An OpenFlow switch has a flow table, and each entry consists of header fields that identify the incoming flow, flow action, and statistical information (see Figure 1, table on the right).

When a packet enters the switch, its fields are compared to the table header. If they match, the corresponding actions occur.

Unlike traditional networks, software-defined networks allow: [3][14]

1. The controller to update the switch's flow table with execution time rules
2. The controller to request switch traffic statistics
3. Flow data paths to change at runtime

These characteristics make the data network more flexible, thus facilitating the implementation of those strategies which make it safer and more dynamic[15].

## 2.2 STATISTICAL INFORMATION

OpenFlow switches send statistical information to the controller when required: by interface, flow, queue, and table. Interface statistics are chosen, so as to illustrate the effectivity of the wavelet transform. The information that the OpenFlow switch sends to the controller includes received and transmitted packets, received and transmitted bytes, received and transmitted drops, and received and transmitted errors.

The information exchange between the OpenFlow switch and controller occurs with two messages: *statistics requests* are used by the controller to request statistical information be sent to the switch, and *statistics reply* is sent by the switch to the controller, in response to a request. The algorithm obtains statistical information from the switches, at 10s-time intervals. The controller sends a request message to all connected switches each time that the temporizer changes. After the controller receives answers from the switches, it chooses the ports associated with the servers, and sends that information to the anomaly detector.

## 3. MONITORING REGISTER CHARACTHERIZATION

Based on the hypothesis that the performance register of a node on a data network, under normal conditions, presents little variation (low frequencies), and that a relatively abrupt or atypical change in the behavior of a node would result in the appearance of high frequencies, it was decided to implement wavelet transform, which would provide information about spatiality and frequency, and be proportional to the wavelet changes in the transform (the k factor in Equation 1). Depending of this factor, the wavelet (($\psi$ t)) either shrinks or dilates [7]. When the analyzed register has low frequencies, a dilated wavelet will allow for the obtention of coefficients of high value, thus indicating the presence of low frequencies. When the register has high frequencies, a contracted wavelet allows for the obtention of high-value coefficients, indicating the presence of high frequencies. The more similar the wavelet to the network activity register form, the higher the wavelet coefficients at a given time [1].

$$W(d,k) = \int_{-\alpha}^{\alpha} x(t) \frac{1}{\sqrt{|k|}} \psi(\frac{t-d}{k}) dt \qquad (1)$$

Equation 1. Wavelet transform.

At the discrete level, the transform is implemented using a bank of filters, low-pass filters perform a process similar to that performed in the continuous transform, in which having a high scaling factor (low frequency extraction) and high pass filters have the same effect as calculating the transform with a small-scale factor (high frequency extraction). Since the goal was to reduce the number of values in the register, the "wavelet packet" method, illustrated in Figure 2, was employed.

The implementation consists of the application of a bank of filters, one step low (LP) and another step high (HP), whose coefficients are determined by the base wavelet. In this case, the register (i) was filtered by performing the convolution operation, thus obtaining a register (i + 1) with a larger number of samples, compared to the signal of the transformation tree's upper level. This necessitated subsampling, in order to reduce the number of samples by half, each time that the filters were applied. *Thus, in the third level of the tree, a signal of 256 samples is obtained, when the original signal initiated the process with 1,024 samples.* The decision to continue applying the discrete transform was determined by the energy level of the filtered signal.

Figure 3 shows the expansion of the tree, via the LP branch, which implies that the energy of the LP register (i + 1) signal is greater than that of the HP (i + 1).

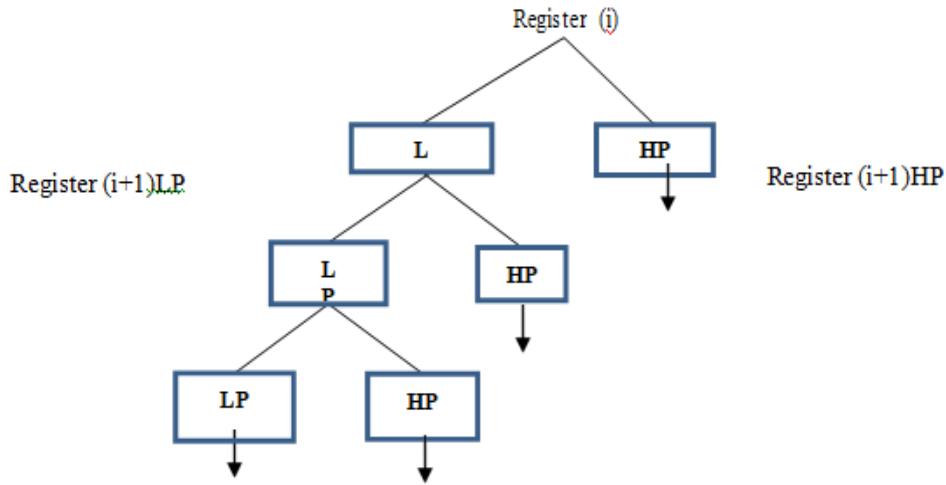

Figure 2. Descompositon tree. Discrete wavelet transform.

## 4. ANOMALY DETECTION METHOD

The detector was implemented in order to determine the degree of reliability of the reduction process. A correct reduction must keep detailed information that permits the performance of all control actions which correspond to data networks.

A Gaussian model was utilized to detect anomalous conditions. The features considered for the model were received and transmitted bytes, which were obtained from the statistical information that the switch sent to the controller. Initially, under normal conditions, an unlabeled dataset was formed: $\{x^{(1)}, x^{(2)}\}$, the first vector, the received bytes, and the second vector, and transmitted bytes.

A Gaussian distribution was fitted into said dataset, and the values with very low probability were considered to be anomalies.

The Gaussian distribution is given by:

$$p(x;\mu,\sigma) = \frac{1}{\sqrt{2\pi\sigma^2}} e^{-\frac{(x-\mu)^2}{2\sigma^2}}$$

(2)

Equation 2. Gaussian distribution.

σ is the variance and μ is the mean. Where the mean and variance are given by:

$$\mu_i = \frac{1}{m}\sum_{j=1}^{m} x_j^i \qquad \sigma_i^2 = \frac{1}{m}\sum_{j=1}^{m}(x_j^i - \mu_i)^2$$

(3)

Equation 3. Gaussian distribution mean and variance.

The algorithm took the initial dataset, and calculated a threshold, so as to determine the lowest probability that should be considered by the anomaly detector to be the limit probability, with which to decide the normality or anomaly of input data.

## 5. EXPERIMENTAL RESULTS

As the basic objective of this investigation was to determine the reliability of the application of wavelet transform, in order to reduce the size of the statistical records obtained by the controller in a software-defined network. This was implemented in a network in the "mininet" simulation environment [11] with controllers programmed in Python. The simulation began with the transmission of a video. "OpenFlow" switches sent statistical information, every 10 seconds, to the controller. Information associated with server interfaces was stored in a file. The parameters selected for performance of the compression process were as follows: bytes transmitted and received during a simulation time of 42 minutes. The filter bank "wavelet" was applied to the 256-byte sequences transmitted (see Figure 3).

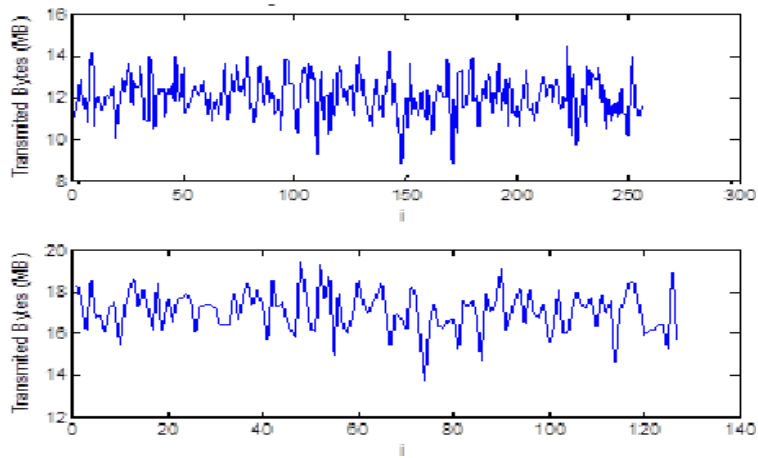

Figure 3. Register reduction with wavelet transform.

The upper graph represents the original register, and shows 256 values of transmitted bytes. The lower graph represents the reduced register, with 128 values, and shows an adequate approximation to the waveform, as well as a reduction of 50% of the original register.

The synthesis process was implemented in order to determine the effectiveness of the application of wavelet transform for this type of data. The inverse wavelet transform was applied to the filter

93

bank output. Figure 4 shows the similarities between the two registers.

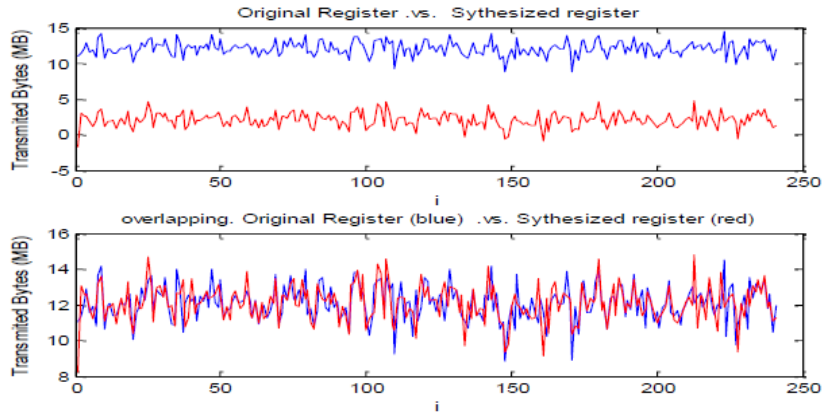

Figure 4. Comparison between the original register (blue) and the wavelet synthesized register (red).

A Gaussian anomaly detector was applied to both the original and synthesized registers, so as to determine the effectiveness of the wavelet transform in the conservation of atypical data see Figure 5. Atypical values are marked with red circles.

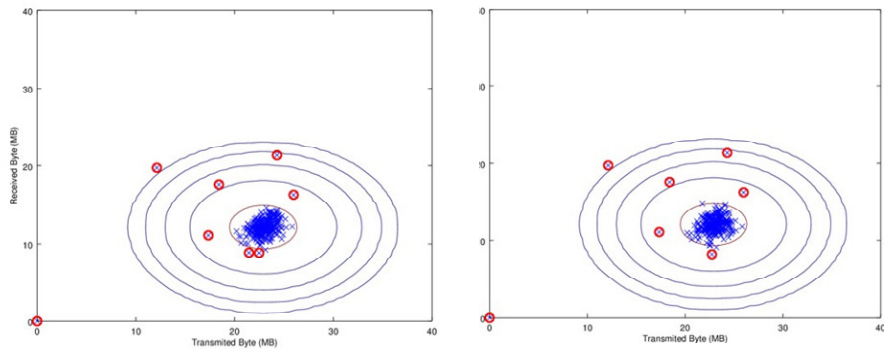

Figure 5. Anomaly detector result. Left, original data. Right, synthetized register.

The conservation of atypical behavior may be observed in Figure. 6.

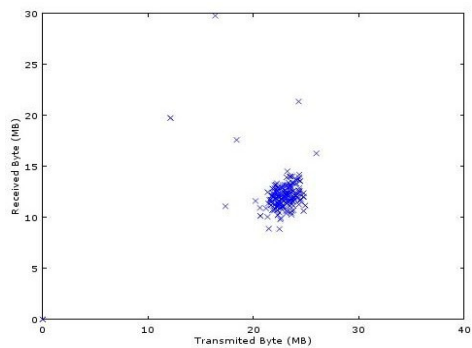

Figure 6. Relationships between atypical points in the reduced register.

This final result verified that the detection of anomalies is possible with a shorter register than the original. This generates lower costs for computing processes and data storage.

## 5. CONCLUSIONS

The reduction algorithm has a compression factor of at least a 50%. It *retains the atypical behavior* of statistical data generated by the OpenFlow switch, which reduces the execution time of the anomaly detector.

The results of this study open the door for analysis of the effectiveness of other wavelet bases in the reduction of the length of statistical parameters in Software-defined Networks. Given that this investigation only used the "daubechies" wavelet, the evaluation of other base wavelets might lead to improved reduction rates.

International Journal of Computer Science & Information Technology (IJCSIT) Vol 11, No 2, April 2019

**AUTHORS**

**Luz A. Aristizábal Q.** is a professor in the Department of Computing in the Faculty of Management at the Universidad Nacional de Colombia. She received her Meng. in Physical Instrumentation from the Universidad Tecnológica de Pereira in 2009, her degree in Data Network Specialization from la Universidad del Valle in 1991, and her degree in Engineering Systems from la Universidad Autónoma in 1989. Her research focuses on aspects of computer and data networks, including network simulators, signal processing, and network paradigms.

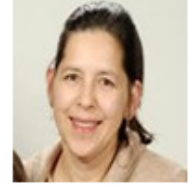

**Nicolás Toro G**. is a professor in the Department of Electricity, Electronics, and Computing. He received his PhD in Engineering-Automation and a bachelor's degree in Electrical Engineering from the Universidad Nacional de Colombia in 2013 and 1983, respectively, and his master's degree in Automated Production Systems from the Universidad Tecnológica de Pereira in 1992. His research focuses on many aspects of industrial automation, including network design, measurement, and analysis.

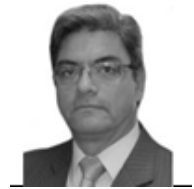